\documentclass{article}
\usepackage{graphicx}
\usepackage{amsthm,amsmath,amsfonts,amssymb}
\usepackage{natbib}
\usepackage[colorlinks,citecolor=blue,urlcolor=blue]{hyperref}
\usepackage{multirow}
\usepackage{authblk}
\usepackage{orcidlink}

\title{Monitoring road infrastructures from satellite images in Greater Maputo: an object-oriented classification approach\footnote{\textit{This is an original manuscript of an article published by Springer in  Statistical Methods \& Applications on 23 December 2024, available at: \url{https://doi.org/10.1007/s10260-024-00772-y}.}}}

\author[1]{Arianna Burzacchi \orcidlink{0000-0001-8284-4909}}
\author[1,2]{Matteo Landrò}
\author[1]{Simone Vantini \orcidlink{0000-0001-8255-5306}}

\affil[1]{MOX Laboratory, Department of Mathematics, Politecnico di Milano, Piazza Leonardo da Vinci 2, 20133, Milano, Italy}
\affil[2]{SAS Institute Inc., 100 SAS Campus Drive, Cary, 27513, NC, USA}

\date{}

\begin{document}

\maketitle

\begin{abstract}
The information about pavement surface type is rarely available in road network databases of developing countries although it represents a cornerstone of the design of efficient mobility systems. This research develops an automatic classification algorithm for road pavement which makes use of satellite images to recognize road segment as \textit{paved} or \textit{unpaved}. The proposed methodology is based on an object-oriented approach, so that each road is classified by looking at the distribution of its pixels in the RGB space. The proposed approach is proven to be accurate, inexpensive, and can be straightforwardly extended from the single case study of Greater Maputo to other cities.
\end{abstract}

            Maputo,
            Satellite images.

\section{Introduction}
\label{sec:intro}

Road infrastructure conditions profoundly affect transport operations \citep{}, and route planning and maintenance could be efficiently managed given a proper classification of road pavement types \citep{Riid2020}. Nevertheless, information on road pavement type is rarely available in developing countries. The OpenStreetMap (OSM) road network of the Greater Maputo Area in Mozambique, for instance, comprises of approximately $12.9$ thousand km of roads, but just for $5$\% of the road network the pavement type is known. The study fits in this context with the goal of developing an automatic classification method for road pavement surface.

This work develops within the Safari Njema project of Politecnico di Milano, winner of Polisocial Award in 2018 and consequently launched in March 2019. The project uses analytical tools to propose strategic solutions for paratransit mobility in African cities. Today, public transit of Sub-Saharan cities is supported by informal transportation services, which account for between $50$-$98$\% of passenger trips and hence are a resource to be integrated into the mobility system \citep{Behrens2016}. Safari Njema project aims at rationalizing paratransit mobility in Sub-Saharan developing countries and providing efficient solutions for redesigning the current mobility offer in more effective and safer plans \citep{web-SAFARI}.

In this paper, we propose an innovative method for automatic road pavement classification from road satellite images. By means of new methodologies and using open-source software and data, roads of unknown surface are labeled as paved or otherwise unpaved. The analysis is conducted in view of low costs and high scalability and could be easily extended from the single case study of Maputo to have a meaningful impact on the (re)design of mobility systems in Sub-Saharan cities.

Research on road pavement mostly develops in the fields of remote sensing for road extraction \citep[e.g.][]{Abdollahi2020} and pavement distress detection \citep[e.g.][]{Ragnoli2018}. To the best of our knowledge, just a few studies explicitly address road pavement type classification. Works in this domain make use of data from vibration and acoustic sensors to classify the terrain type \citep[e.g.][]{Li2019, Paulo2018}. Other studies exploit image-based classification for road surfaces. In \citet{Riid2020}, panoramic images of roads are recorded, and their type is predicted by a convolutional neural network. \citet{Slavkovikj2014} make use of Google Street View images for unsupervised feature extraction and SVM classification. Starting from Google Street View imagery, \citet{Marianingsih2019} propose to use GLCM to describe texture features and a combination of $k$-NN and Na\"{i}ve Bayes to characterize road surfaces.

This research develops in the realm of Object-Oriented Data Analysis \\(OODA) \citep{Marron2014, Marron2021}. The proposed method works with Google Earth satellite images of the Greater Maputo Area and map features exported from the OSM road network. Their information is the input of an object-oriented supervised classification algorithm which, starting from roads of known surface type, predicts pavement surface of the remaining $95$\% of roads. The statistical unit for classification is the set of pixels representing the street surface in the image. For each cloud of street pixels of unknown road pavement, the classification is made by considering the pavement type of the most similar clouds of street pixels, namely the ones with a smaller distance in the embedding color space.

This paper is organized as follows. Section \ref{sec:data} presents the dataset, describing methods from data extraction and pre-processing for pixel selection. In section \ref{sec:analysis}, the classification algorithm is introduced. Firstly, the focus is on the choice of the best mathematical embedding for the object-oriented approach. Then, the algorithm is trained and tuned to minimize the cost of its application in real-world situations. Finally, conclusions are discussed in section \ref{sec:conclusion}, together with suggestions for further developments.

\section{From satellite images to data points} \label{sec:data}

\begin{figure}
\centering
\includegraphics[scale=0.2]{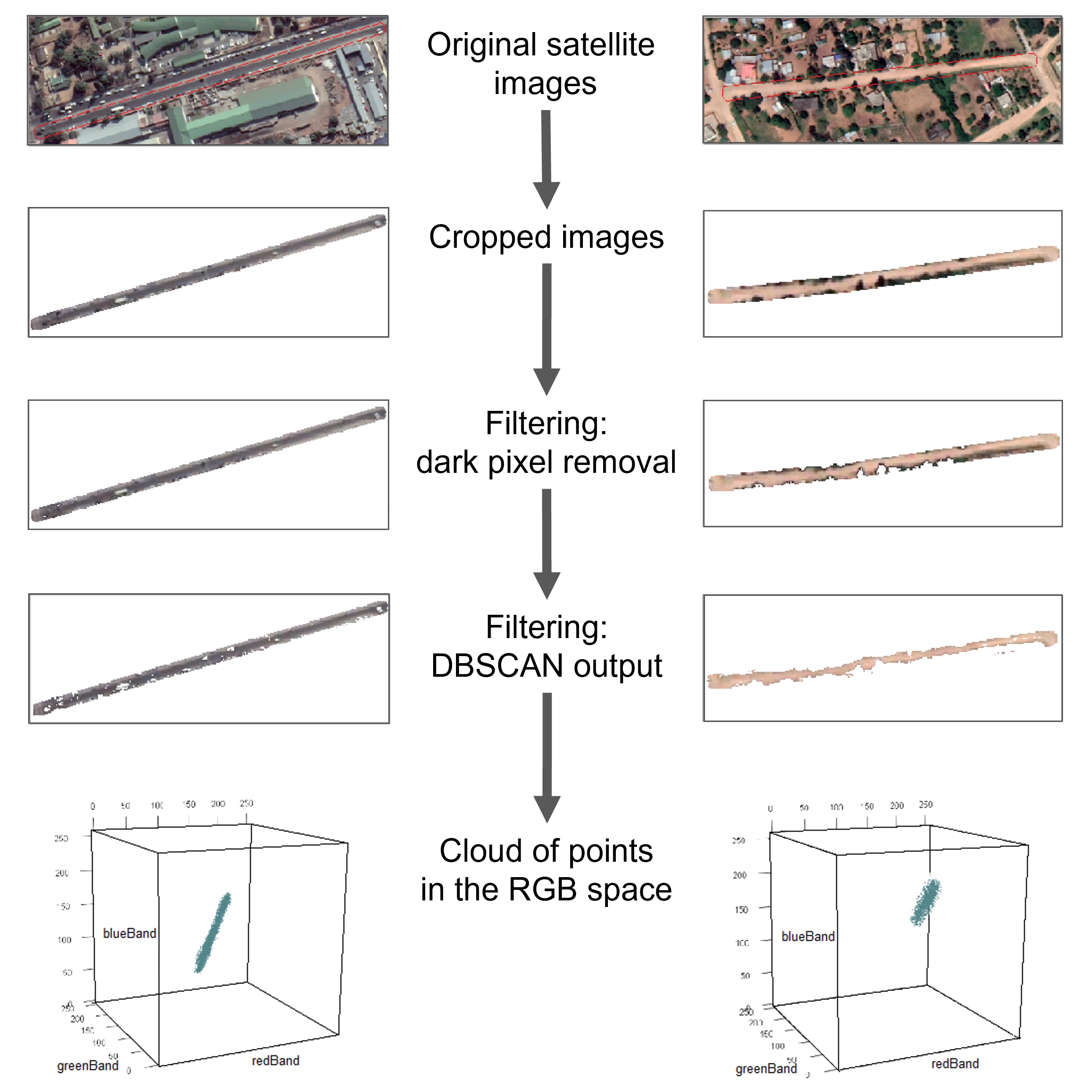}
\caption{Image processing pipeline: from raw satellite images to data points for the object-oriented classification}
\label{fig:DBSCAN}
\end{figure}

Two main data sources were used in the analysis: the OSM road network and its information, and Google Earth satellite images.

The OSM road network \citep{OSM} was split into $53\,240$ segments, one for each stretch of the road, with a length of between $50\,\text{m}$ and $550\,\text{m}$. Shorter paths were not considered, and larger ones were split into sub-parts. Each street segment is described by its OSM map features, some available for each segment, like GPS coordinates and street type, while some present for just a minority of segments and unknown for others. It is the case of the road surface type. The pavement type indicates whether a road is \textit{paved} or \textit{unpaved}. More specific tags such as \textit{asphalt} \citep[see][]{wiki-OSM}, were merged with the main two classes. Among all, only less than $5$\% of roads have a known pavement composition. Specifically, $732$ streets are paved and $1\,826$ unpaved.

High-resolution satellite images were collected from Google Earth in April 2020 \citep{google-earth-imagery}. They cover the Greater Maputo Area, including both the city of Maputo and the neighboring districts of Marracuene, Matola, and Boane, with a total surface of \mbox{$1\,568\,\text{km}^2$}. Satellite data were clipped by means of QGIS software 
using the OSM road network with a \mbox{$7\,\text{m}$} wide buffer around the road polyline. The buffer ensures that the actual road is captured in the image despite possible spatial inaccuracies of OSM polyline positioning. $53\,240$ raster images were hence obtained, one for each segment of the OSM road network, with three color bands (RGB) and pixel resolution of $1.1$x$1.1\, \text{m}^2$.

Both due to the $7\,\text{m}$ buffer with which the images were extracted, and the natural covering elements of the streets, image data not only represent roads, but also vehicles, vegetation, and buildings. For this reason, a pre-processing phase was needed to remove uninformative pixels from each image. Figure \ref{fig:DBSCAN} outlines the main steps of the process: from the original satellite images, cropped with the $7\,\text{m}$ wide buffer, filtering was performed to obtain clouds of points associated with road pavements.

The pixel filtering was in two steps. In the first step, pixels associated with the canopy were removed simply using a threshold on their darkness, i.e., all pixels whose Euclidean norm in the RGB space was less than or equal to a certain threshold ($t=90$) were not considered. This simple rule was proven to be effective and outperformed more sophisticated methods.

Street pixels were subsequently identified with a data-driven filtering approach by means of DBSCAN. Each image shows the main road path, whose pixels have almost the same color and are distributed around the diagonal of the RGB cube. Pixels of buildings, vehicles, and additional objects, when present, form small disjoint clusters connected to the main pixel cloud by some background noise. Based on these observations, the density-based method \mbox{DBSCAN} was chosen among other clustering algorithms to identify street pixels. Firstly proposed by \citet{Ester1996}, the \mbox{DBSCAN} (Density-Based Spatial Clustering of Application with Noise) algorithm is widely used to discover clusters of any arbitrary shape and size in databases containing noise and outliers by integrating spatial connectivity and color similarity. \mbox{DBSCAN} works with two user-specified input parameters: $\varepsilon$, as the neighborhood radius, and \textit{MinPts}, as the minimum number of points in the neighborhood. Parameter tuning is one of the limitations of the methodology and some parameter-free implementations of DBSCAN were proposed \citep[see, for instance,][]{Ester1996, Kurmalla2016, Hou2016}. However, in this instance, visual inspection of data and cluster shape was preferred to define a suitable criterion for parameter choice.

A common feature among all images is the distribution of street pixels around the diagonal of the RGB cube, as previously mentioned. Nevertheless, they differ by the density of points in the space. The number of pixels per image ranges from $200$ to $5\,000$, generating clouds of different sizes and densities in the RGB space. Thus, to get a proper street pixel extraction, $\varepsilon$ and \textit{MinPts} should differ for each image.

\begin{figure}[tb]
       \centering
       \includegraphics[scale=0.13]{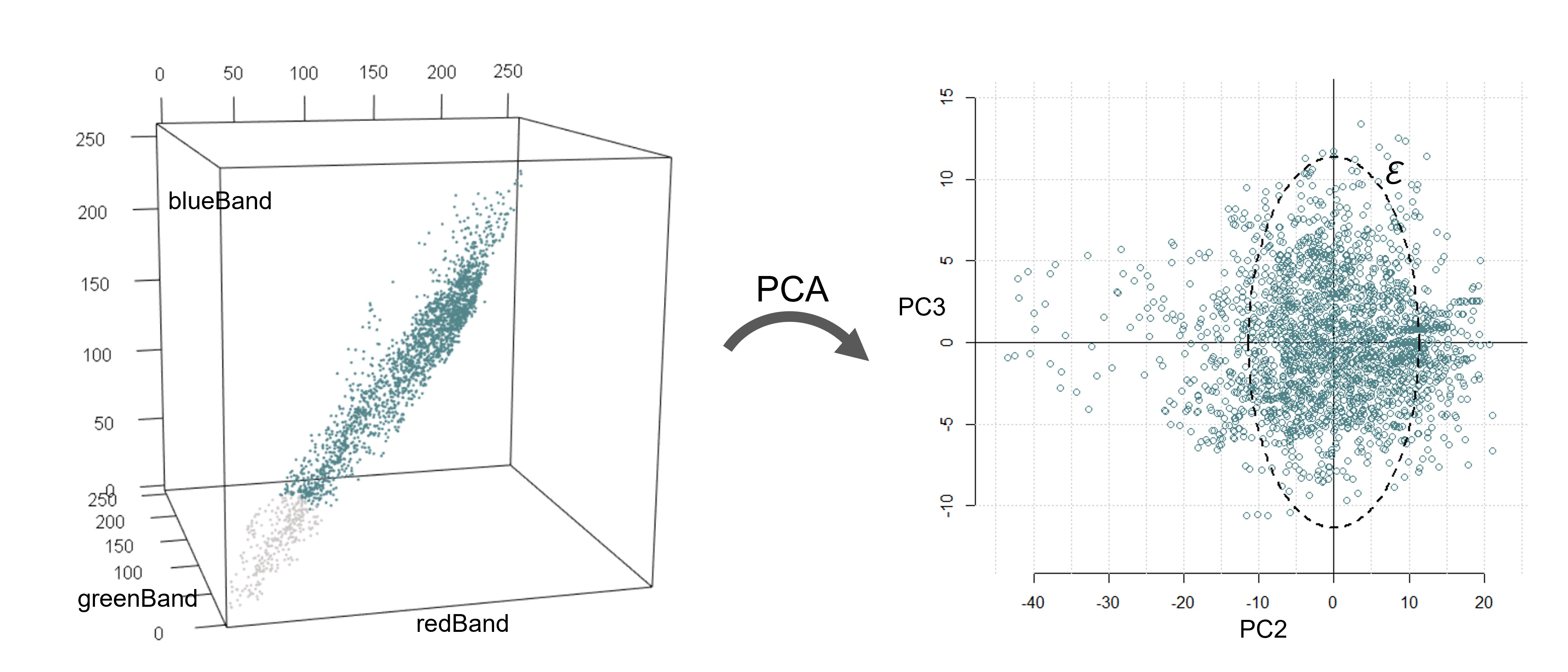}
       \caption{Cloud of points in the RGB space (on the left) and its projection in the plane of second and third principal components (on the right)}
       \label{fig:img_2}
    \end{figure}

\paragraph*{Tuning of the parameter $\varepsilon$} The choice of $\varepsilon$ considers the color variability of the image, i.e., the Euclidean distance of points from the diagonal of the RGB cube. Specifically, $\varepsilon$ was chosen as the $0.75$ quantile of such a distance. The definition of $\varepsilon$ as the $0.75$ quantile of distances was preferred to the maximum distance to exclude highly saturated outliers from the cluster of street pixels. 
	One can observe that, given the shape of pixel distributions, the diagonal is the maximum variance direction in the data. Hence, for each image, the first principal component of PCA almost coincides with the diagonal. This fact allowed to approximate the actual distance between point and diagonal in the $3$-dimensional RGB space, as the distance between point and origin in the $2$-dimensional space of second and third principal components (PC\textsubscript{2} and PC\textsubscript{3}). Accordingly, $\varepsilon$ was chosen as the $0.75$ quantile of Euclidean norm of the projection of points in the plane whose axes are PC\textsubscript{2} and PC\textsubscript{3}. Figure \ref{fig:img_2} shows the points in the original RGB cube and their projection on the plane of PC\textsubscript{2} and PC\textsubscript{3}.
	
\paragraph*{Tuning of the parameter MinPts} 
    The selection criterion for \textit{MinPts} exploits the relationship between the density of the whole cloud of points, $d_{cyl}$, and the minimum density in the \mbox{$\varepsilon$-neighborhood} of each core point, $d_{neig}$. We claim that the \mbox{$\varepsilon$-neighborhood} of a core point is always denser than the cloud of street pixels. The statement is supported by the intuition that the street cluster is expected to have a higher concentration of points than the whole cloud with uniformly distributed points.
	To compute $d_{cyl}$ the cloud of points is approximated as a cylinder of height $h$ and radius $\varepsilon$ where $75$\% of the pixels are uniformly distributed. Since dark vegetation pixels had already been removed, $h=diag_{\text{\tiny{RGB}}}-t$.
	\begin{equation*}
	    d_{cyl} = d_{cyl}(\varepsilon) = 0.75\: n_{pts}/(h \: \pi \varepsilon^{2})
    \end{equation*}

    $d_{neig}$ is the minimum density of the \textit{$\varepsilon$}-neighborhood of a point required to be considered a core point:
    \begin{equation*}
	    d_{neig} = d_{neig} (\varepsilon, MinPts) = MinPts/(\tfrac{4}{3} \pi \varepsilon^3)
    \end{equation*} 

    Given a fixed $a>1$, 
    \begin{equation}\label{eq:DneigDcyl} 
	    d_{neig} (\varepsilon, MinPts) = a \: d_{cyl}(\varepsilon)
    \end{equation}

    Solving the equality in (\ref{eq:DneigDcyl}), the definition of \textit{MinPts} is easily found: 
    \begin{equation}\label{eq:MinPts}
	    MinPts = MinPts(\varepsilon) = a \: n_{pts}\: \varepsilon /h
        \end{equation}

    Note that \textit{MinPts} depends on $\varepsilon$, so \textit{MinPts} must be computed after $\varepsilon$ is found. There is no indication in the literature about the choice of $a$, since this whole parameter-tuning procedure is inspired by the data under examination. Therefore, the constant $a$ was chosen empirically to be $\tfrac{4}{3}$ since better results were obtained.

Using $\varepsilon$ and \textit{MinPts} as above, DBSCAN was performed for each image to detect the cluster of significant street pixels. If more than one cluster were found, the most numerous one was selected. All computations were implemented in R 
and made use of pre-implemented functions of R libraries \citep{pack-fpc, pack-raster}. 

\begin{figure}[tb]
   \centering
   \includegraphics[scale=0.13]{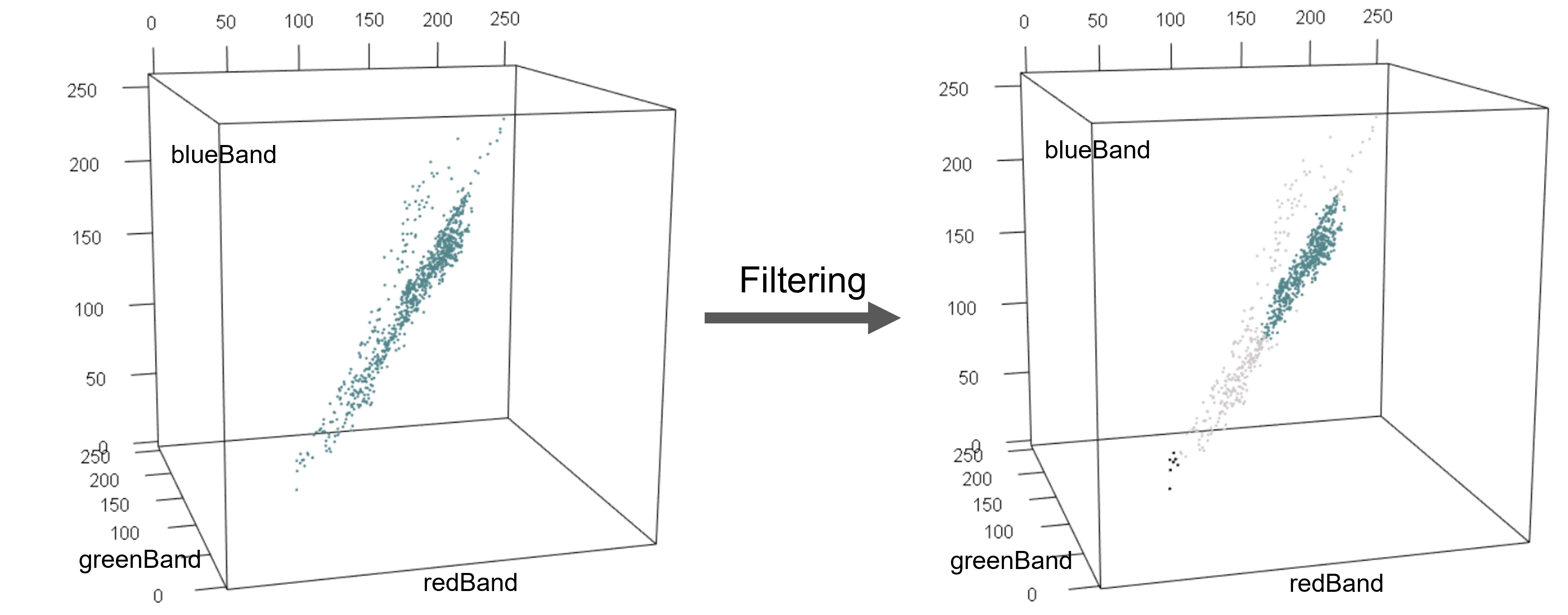}
   \caption{Cloud of points in the RGB space of a specific road satellite image. The figure on the left shows all the points of the satellite image, while the figure on the right represents the points related to the road surface (in blue), dark pixels (in black), and other pixels (in grey)}
   \label{fig:img_3}
 \end{figure}

In conclusion, starting from the cropped satellite images, by means of a two-step filtering process, we were provided with the set of pixels related to the road surface for each image. Figure \ref{fig:img_3} shows both the input cloud of points of a satellite image and the output of this phase, where street pixels are highlighted among others. Those street pixels in the RGB space were the data points for subsequent analysis.

\section{Data analysis} \label{sec:analysis}

\subsection{Object-oriented classification} \label{subsec:k-NN}

Once that the street pixels were extracted from each image, an algorithm for road pavement classification was built. $2\,558$ street segments of known pavement surface were used to train a $k$-NN classification algorithm and subsequently label unknown pavement roadways as paved or unpaved.

The study takes an innovative object-oriented approach to this task within the domain of OODA. Groups of pixels are modeled mathematically as clouds of points in the color space. Then, classification is achieved by means of the $k$-nearest neighbors method with clouds of street pixels as input instances. The main rationale is that streets whose pixels are similarly distributed in the color space have also the same type of road surface. So, given a cloud of street pixels, $k$-NN assigns a specific label, either paved or unpaved, depending on the labels of the $k$ most similar clouds of street pixels. More precisely, each street pixel cloud was treated as a random sample from a multivariate distribution in the color space. Given a similarity measure among probability distributions, the $k$ closest distributions were found and used for classification.

Four metrics between distributions were taken into account. A first trial was made with the Euclidean distance between cloud Tukey medians \citep{Tukey1974, Rousseeuw1998}. The information of the whole cloud of points was summarized in its single most central point, the Tukey median, and distances were computed in the Euclidean space of those representative points. Considering instead the whole distribution of points defined on a measurable space, there were computed the Hausdorff distance between clouds \citep{Huttenlocher1993}, the Energy distance between clouds \citep{Szekely2004}, and the Wasserstein distance between clouds \citep{Panaretos2019}. They were then compared by evaluating the misclassification error rate (MER) and computational time of $k$-NN on the test set. To speed up computations, $n\!=\!150$ points were randomly sampled from the street pixels set of each image to represent pixel distributions (see appendix \ref{app:N-choice} for additional details on the choice of $n$). Energy distance was preferred to all the others both for low MER and computational time, as shown in table \ref{tab:metrics}, and for its simple implementation. It is defined as follows:

\small{
\begin{equation*}\displaystyle
	d_{e}(A,B) = \tfrac{n_a n_b}{n_a+n_b} \left[ \tfrac{2}{n_a n_b} \sum_{a\in A,b \in B} d(a,b) 
	-  \tfrac{1}{n_a^2}\sum_{a,a' \in A} d(a,a') - \tfrac{1}{n_b^2} \sum_{b,b'\in B} d(b,b') \right]
\end{equation*}
}
\\ 
where $A, B$ are finite subsets of a metric space $(M,d)$ (i.e., the two clouds of pixels), $|A| = n_a$, $|B| = n_b$ (i.e., the two cloud sizes). The ground distance $d$ was chosen to be the Euclidean distance in the RGB space: \\
\mbox{$d(a,b) = \sqrt{(R_a - R_b)^2+ (G_a - G_b)^2 + (B_a - B_b)^2}$}. Further information about the metric definition and computation is contained in appendix \ref{app:distances}.

\begin{table}[tb]
    \centering
    \caption{MER and computational time when different metrics for distributions are used in the object-oriented $k$-NN}  
    \label{tab:metrics}
    \begin{tabular}{lcc}
        \hline
        \textbf{Distance} & \textbf{MER} & \textbf{Computational time} \\
        \hline
        Euclidean between Tukey medians & $0.1943$ & $1.66\,\text{s}$ \\
        Hausdorff & $0.1369$ & $8.11\,\text{s}$ \\
        Energy & $0.1082$ & $5.67\,\text{s}$ \\
        Wasserstein & $0.1108$ & $21.40\,\text{s}$\\
        \hline  
    \end{tabular}     
\end{table}

It was then investigated whether the RGB color space was convenient for this analysis. The RGB space is commonly considered the base color model and is widely used in many fields, but also other color spaces are defined and might overcome the limitations of RGB. For instance, the R'G'B' space is introduced to model the non-linearity of human color perception, and the HSV space is able to describe the amount of physical quantities through its components' hue, saturation and brightness. Moreover, histogram equalization techniques could be easily applied in the HSV space, rather than in the RGB space, due to the low correlation between components. Therefore, in this analysis, data were transformed in alternative color spaces to test which one was the most suitable for this application. As shown in table \ref{tab:colors}, misclassification error rates are comparable for $k$-NN with Energy distance in RGB, R’G’B’ and HSV spaces. The latter minimizes MER, but its computation is the most time-consuming. Therefore, RGB space was preferred. More details on color spaces are contained in \citet{Plataniotis2013} and appendix \ref{app:colors}. 

\begin{table}[tb]
    \centering
    \caption{MER of $k$-NN classification in different color spaces}
    \label{tab:colors}
    \begin{tabular}{l@{\hskip 50pt}c}
        \hline
        \textbf{Color space} & \textbf{MER}\\
        \hline
        RGB & $0.1082$  \\
        R'G'B' & $0.1108$ \\
        HSV & $0.1056$  \\
        HSV with equalization & $0.1551$ \\
        \hline 
    \end{tabular}
\end{table}

Finally, an object-oriented $k$-NN classification algorithm was built. Street pixel distributions in the RGB space were used as input and their relative distances were measured with the Energy metric. The number of neighbors, $k$, was set equal to $5$ to minimize the misclassification error of 10-fold Cross-Validation over the training set of $70$\% of the known pavement roads.

\subsection{Threshold choice by cost minimization} \label{subsec:costmin}

The $k$-NN classifier measures the frequency of paved and unpaved roads among the $k$ nearest street pixel clouds in the RGB space to choose the output class label. Threshold $f_{up}\in\! [0,1]$ refers to the minimum frequency of paved roads among the $k$ nearest so that a street segment is classified as paved by the algorithm: when the paved road frequency $f$ is greater or equal to $f_{up}$, the road is classified as paved, otherwise as unpaved.

The choice of the classification frequency threshold is typically related to the minimization of MER. The application of this traditional criterion to our problem would have picked a frequency threshold $f_{up}=0.6$, ending up in a model with an accuracy of $89.2$\% on the test set. Nevertheless, the real-world context of application of the classification algorithm is not considered by this criterion. It is instead a standard approach, which does not consider the actual cost of application of the algorithm to real-world situations. The frequency threshold choice was hence inspired by a deeper and more practical reasoning, both introducing the realistic scenario of uncertain output of the algorithm and then quantifying the estimated impact of misclassification errors and uncertainties.

First of all, the classification algorithm was allowed to retain uncertain decisions. It is a widely used approach in pattern recognition applications \citep[see, for instance,][]{Chow1970, Pillai2013}. Especially when the accuracy is not satisfactory and misclassification costs can be overcome with the use of much cheaper manual labeling, the option to avoid unreliable classification is introduced to prevent excessive misclassification and to leave the decision to human operators.

The approach is applied through the introduction of two distinct thresholds: $f_{p}$, as the minimum paved frequency so that the street is classified as paved, and $f_{u}$, indicating the maximum paved frequency to identify an unpaved road, with $0\leq f_{u}\leq f_{p}\leq1$. Under this circumstance, the classifier is allowed to be uncertain about a particular road which has “neither enough paved neighbors nor unpaved neighbors” in the RGB space (i.e., $f_{u} \leq f \leq f_{p}$), and so is classified as \textit{uncertain}.

The definition of both classification rules follows, showing both cases of single threshold $f_{up}$, when uncertainty is not permitted, and pair of thresholds $\{f_{u}, f_{p}\}$:

\begin{alignat*}{2}
	& \begin{aligned} & \begin{cases} 
		\text{  if } f \geq f_{up} \Rightarrow \text{paved}\\ 
		\text{  if } f < f_{up} \Rightarrow \text{unpaved}\\ 
		\end{cases}
	\end{aligned}
	& \hskip 4em 
	& \begin{aligned} & \begin{cases}
		\text{  if }f > f_{p} \Rightarrow \text{paved}\\ 
		\text{  if }f < f_{u} \Rightarrow \text{unpaved}\\ 
		\text{  if }f_{u} \leq f \leq f_{p} \Rightarrow \text{uncertain}\\
		\end{cases}
	\end{aligned} 
\end{alignat*}

Furthermore, account was taken of the expected impact of the classification outcomes. Every output of the algorithm has different practical consequences when employed to improve the mobility system. Coherently, classification costs were assigned to each misclassification or uncertain scenario.

The choice of cost values focuses on the practical impact of the classification output on single users, in terms of traffic delays, wear and tear of vehicles, and road accidents \citep{Tsunokawa2002}. Firstly, it should be noted that the misclassification cost is not symmetric. Misclassifying unpaved roads as paved could trouble travelers with severe car damage and travel delays, whereas the opposite misclassification error would give a pejorative picture of reality, discouraging rides along alternative routes and increasing traffic congestion. Since the impact is different, there are defined different cost values according to the type of road pavement wrongly classified. The misclassification cost $c_{up}$ measures the effect of wrongly classifying actual unpaved roads as paved and, conversely, $c_{pu}$ represents the impact of misclassifying paved streets as unpaved. An additional cost is associated with the uncertainty of the algorithm when classifying road surfaces. Indeed, doubtful situations would result in the labor-intensive and time-consuming intervention of employees who manually classify uncertain road surfaces, taking over the goal of road classification from the algorithm. The uncertainty cost $c_{d}$ was then defined to quantify how much should be paid in case of doubtful situations.

With the help of domain experts, there were assigned the following unit costs: $c_{up}=2.5$, $c_{pu}=2$ and $c_{d}=1$. The proposed cost values are not to be intended as absolute costs, but properly represent the magnitude of each doubtful or mistaken situation in relation to the others. Regarding the cost of uncertainty, $c_{d}$, it should be smaller than error costs. It is better for a user to be aware of the uncertainty rather than to find unexpected and undesired road conditions. Moreover, both situations would need human inspections to be solved, but while mistakes need to be both discovered and adjusted, doubts just need a correction. Then, $c_{up}$ is said to be greater than $c_{pu}$. The damage caused by the first type of misclassification is considered to be more relevant than that of the second type. More than half of unpaved roads in Sub-Saharan countries are not passable \citep{JICA2010}, and wrongly detouring on unpaved roads could slow down and hence delay travel, or cause severe damage to vehicles. This situation is considered to be more serious than the usual traffic delay on main paved roads, which might be enhanced by the opposite mistake. For these reasons, $c_{d} \leq c_{pu} \leq c_{up}$.

The application of the algorithm has an expected total cost equal to the sum of unit costs weighted by the number of mistakes and doubts of classification. The proposed classification algorithm was hence refined in such a way that, if applied, it would have the expected total cost as low as possible. The $k$-NN classification algorithm with $k=5$ and Energy distance between RGB street pixel clouds was trained and then applied to the remaining test set. Different thresholds were used and brought to different misclassification and uncertainty proportions, and so to different total costs. Then, among all possibilities, the best algorithm was chosen as the one minimizing the total cost. Tested values of the single threshold span between 0 and 1: $f_{up}=i/k$ for $i=0,\dots,k$, with $k=5$. Similarly, all values for pair of thresholds $\{f_{u}, f_{p}\}$ were tried: $f_{p}= i/k$ for $i = 0,\dots,k$, and $f_{u}=j/k$ for $j\leq i$, with $k=5$. Table  \ref{tab:costs} summarizes the estimated total cost of each classifier when using the proposed unit costs. The minimum total cost $c_{tot}=189.0$ is reached by the algorithm with the following classification rule: if $f\in \{0; 0.2\}$ the road is classified as unpaved, if $f=0.4$ the road is classified as uncertain, and if $f \in \{0.6; 0.8; 1\}$ the road is classified as paved.

\begin{table}
    \centering
     \caption{Errors, uncertainty and cost using different classification rules}
     \label{tab:costs} 
    \scriptsize{
	\begin{tabular}{lccccccc}
		\hline
		& {single} & {j=0} &  {j=0.2} &  {j=0.4} &  {j=0.6} &  {j=0.8} &  {j=1} \\ \hline
			
		\multirow[c]{4}{0.7cm}{{i=0}} & N\textsubscript{up} $=553$ & N\textsubscript{up} $=149$ & N\textsubscript{up} $=61$& N\textsubscript{up} $=30$& N\textsubscript{up} $=10$& N\textsubscript{up} $=2$& N\textsubscript{up} $=0$\\
		
		& N\textsubscript{pu} $=0$ & N\textsubscript{pu} $=0$ & N\textsubscript{pu} $=0$& N\textsubscript{pu} $=0$& N\textsubscript{pu} $=0$& N\textsubscript{pu} $=0$& N\textsubscript{pu} $=0$\\
		
		& N\textsubscript{d} $=0$ & N\textsubscript{d} $=420$ & N\textsubscript{d} $=522$& N\textsubscript{d} $=576$& N\textsubscript{d} $=632$& N\textsubscript{d} $=675$& N\textsubscript{d} $=767$\\
		
		& \textbf{c} $=\mathbf{1106.0}$ & \textbf{c} $=\mathbf{718.0}$ & \textbf{c} $=\mathbf{644.0}$ & \textbf{c} $=\mathbf{636.0}$ & \textbf{c} $=\mathbf{652.0}$ & \textbf{c} $=\mathbf{679.0}$ & \textbf{c} $=\mathbf{767.0}$\\ \hline

		\multirow[c]{4}{0.7cm}{{i=0.2}} & N\textsubscript{up} $=149$ && N\textsubscript{up} $=61$& N\textsubscript{up} $=30$& N\textsubscript{up} $=10$& N\textsubscript{up} $=2$& N\textsubscript{up} $=0$\\
		
		& N\textsubscript{pu} $=16$ && N\textsubscript{pu} $=16$& N\textsubscript{pu} $=16$& N\textsubscript{pu} $=16$& N\textsubscript{pu} $=16$& N\textsubscript{pu} $=16$\\
		
		& N\textsubscript{d} $=0$ && N\textsubscript{d} $=102$& N\textsubscript{d} $=156$& N\textsubscript{d} $=212$& N\textsubscript{d} $=255$& N\textsubscript{d} $=347$\\
		
		& \textbf{c} $=\mathbf{338.0}$ && \textbf{c} $=\mathbf{264.0}$ & \textbf{c} $=\mathbf{256.0}$ & \textbf{c} $=\mathbf{272.0}$ & \textbf{c} $=\mathbf{299.0}$ & \textbf{c} $=\mathbf{387.0}$\\ \hline

		\multirow[c]{4}{0.7cm}{{i=0.4}} & N\textsubscript{up} $=61$ &&& N\textsubscript{up} $=30$& N\textsubscript{up} $=10$& N\textsubscript{up} $=2$& N\textsubscript{up} $=0$\\
		
		& N\textsubscript{pu} $=30$ &&& N\textsubscript{pu} $=30$& N\textsubscript{pu} $=30$& N\textsubscript{pu} $=30$& N\textsubscript{pu} $=30$\\
		
		& N\textsubscript{d} $=0$ &&& N\textsubscript{d} $=54$& N\textsubscript{d} $=110$& N\textsubscript{d} $=153$& N\textsubscript{d} $=245$\\
		
		& \textbf{c} $=\mathbf{197.0}$ &&& \textbf{c} $=\mathbf{189.0}$ & \textbf{c} $=\mathbf{205.0}$ & \textbf{c} $=\mathbf{232.0}$ & \textbf{c} $=\mathbf{320.0}$\\ \hline

		\multirow[c]{4}{0.7cm}{{i=0.6}} & N\textsubscript{up} $=30$ &&&& N\textsubscript{up} $=10$& N\textsubscript{up} $=2$& N\textsubscript{up} $=0$\\
		
		& N\textsubscript{pu} $=53$ &&&& N\textsubscript{pu} $=53$& N\textsubscript{pu} $=53$& N\textsubscript{pu} $=53$\\
		
		& N\textsubscript{d} $=0$ &&&& N\textsubscript{d} $=56$& N\textsubscript{d} $=99$& N\textsubscript{d} $=191$\\
		
		& \textbf{c} $=\mathbf{192.5}$ &&&& \textbf{c} $=\mathbf{208.5}$ & \textbf{c} $=\mathbf{235.5}$ & \textbf{c} $=\mathbf{323.5}$\\ \hline

		\multirow[c]{4}{0.7cm}{{i=0.8}} & N\textsubscript{up} $=10$ &&&&& N\textsubscript{up} $=2$& N\textsubscript{up} $=0$\\
		
		& N\textsubscript{pu} $=89$ &&&&& N\textsubscript{pu} $=89$& N\textsubscript{pu} $=89$\\
		
		& N\textsubscript{d} $=0$ &&&&& N\textsubscript{d} $=43$& N\textsubscript{d} $=135$\\
		
		& \textbf{c} $=\mathbf{242.5}$ &&&&& \textbf{c} $=\mathbf{269.5}$ & \textbf{c} $=\mathbf{357.5}$\\ \hline

	    \multirow[c]{4}{0.7cm}{{i=1}} & N\textsubscript{up} $=2$ &&&&&& N\textsubscript{up} $=0$\\
		
		& N\textsubscript{pu} $=124$ &&&&&& N\textsubscript{pu} $=124$\\
		
		& N\textsubscript{d} $=0$ &&&&&& N\textsubscript{d} $=92$\\
		
		& \textbf{c} $=\mathbf{314.0}$ &&&&&& \textbf{c} $=\mathbf{402.0}$\\ \hline
	\end{tabular}}
\end{table}

The optimal $k$-NN classifier correctly classifies $85.1$\% of the road surfaces and predicts as \textit{uncertain} $7.0$\% of the street segments of the test set. Confusion table \ref{tab:confusion1} shows that uncertainty and mistakes are almost equally distributed between paved and unpaved segments.

\begin{table}
    \centering
    \caption{Confusion table of the optimal $k$-NN classifier from satellite images over the test set, with $\text{k}\!=\!5$ and pair of thresholds $f_{p}\!=\!f_{u}\!=\!0.4$}
    \label{tab:confusion1}
    \begin{tabular}{p{0.8cm}|p{1.5cm}|p{1.5cm}|p{1.5cm}|p{1.5cm}|}
        \multicolumn{2}{c}{}&\multicolumn{3}{c}{\textit{Predicted}}\\
	\cline{3-5}
	\multicolumn{2}{c|}{}&paved &unpaved &uncertain\\
	\cline{2-5}
	\textit{True} &paved & $161$ & $30$ & $23$ \\
	\cline{2-5} &unpaved & $30$ & $492$ & $31$ \\
	\cline{2-5}
    \end{tabular}
\end{table}

\subsection{Classification with images and street type} \label{subsec:type}

OSM provides an internal classification of roads according to their function and relevance in the road network. Most roads in Maputo serve as access to properties (\textit{residential}, $84.6$\%) and low-level connection of the grid network (\textit{unclassified}, $8.6$\%). Just a small portion of roads are employed as urban and interurban connections: \textit{primary} ($1.8$\%), \textit{secondary} ($1.1$\%), and \textit{tertiary} ($2.6$\%). Finally, \textit{footways} ($1.3$\%) map minor pathways, mainly used by pedestrians. For a detailed description of type categories, see \citet{wiki-OSM}.

Streets of the same type are theoretically characterized by more similar infrastructure conditions and pavement typology. Type information was hence exploited to improve the classification of road pavement surface. It was included in the method with a slight modification of the neighbor research. Instead of looking for the most similar street images \textit{in general}, the extended algorithm was imposed to consider only the $k$ nearest streets \textit{of the same typology} in the RGB space. Appendix \ref{app:type} explains formal details about this study, from parameter choice to additional investigation on road types. The classification accuracy improves reaching $94.5\%$ of correctly classified road surfaces. Over $767$ streets of the test set, only $28$ are misclassified paved and $14$ are misclassified unpaved, resulting in a MER of $0.0548$ and no uncertain roads.

\begin{figure}[ht!]
    \centering
    \includegraphics[scale=0.25]{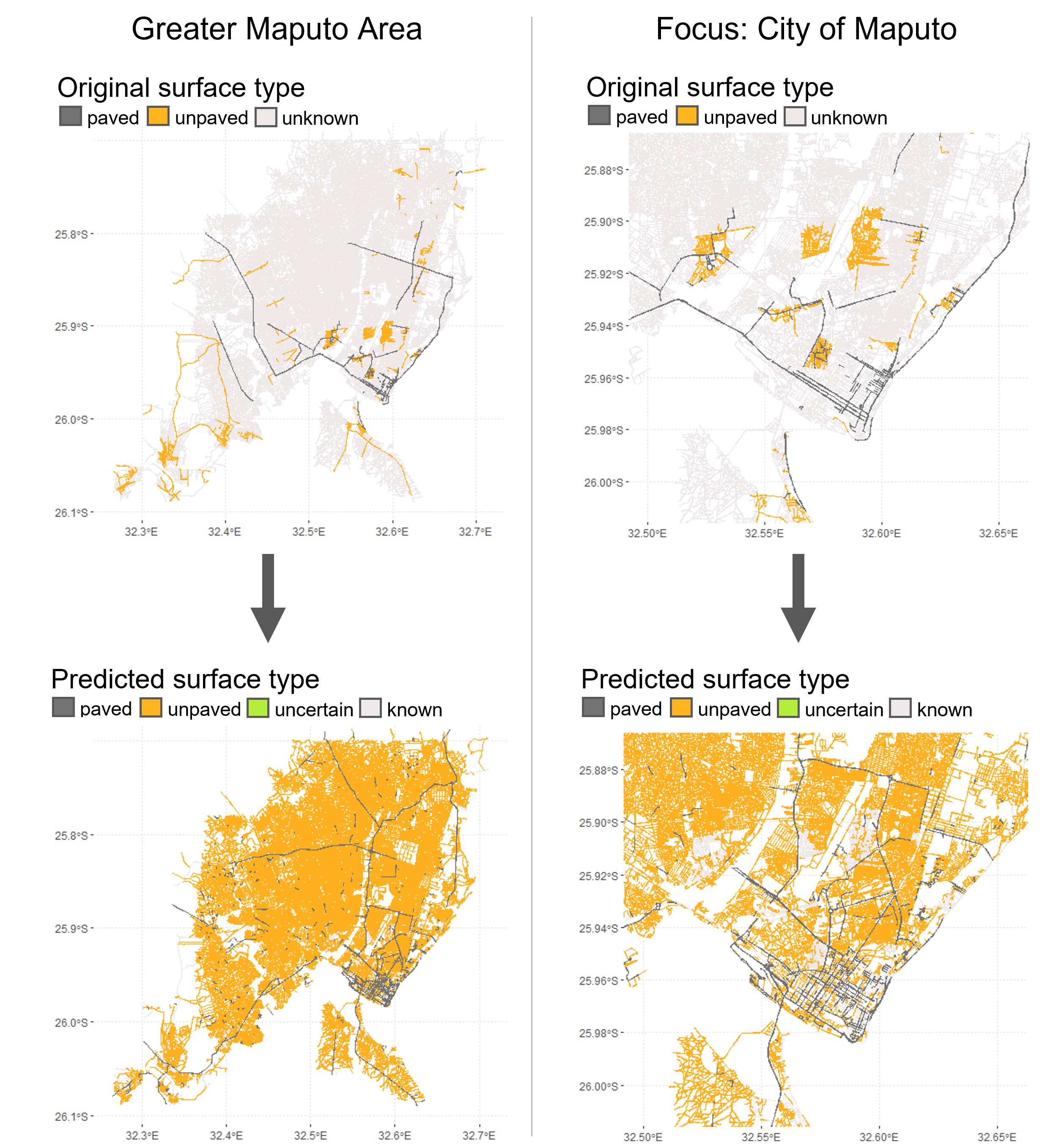}
    \caption{Maputo road network colored by known pavement type (top images) and predicted pavement type (bottom images). The road network of the whole Greater Maputo Area is represented by the figures on the left, and a zoom on the city of Maputo is shown on the right}
    \label{fig:Maputo}
\end{figure}

The mixed approach which combines color pixels data and street type information was applied to predict unknown road pavements. With the described color space, distance and parameters, $k$-NN algorithm was trained with $2\,558$ street data of known pavement and gave a highly accurate picture of the predicted road network pavement in Maputo, as shown in figure \ref{fig:Maputo}.

\section{Conclusion} \label{sec:conclusion}

In this paper, a new method for the classification of road pavement surface is proposed. An object-oriented $k$-NN classification algorithm uses Google Earth satellite images and, when available, OSM street type information, to obtain an accurate prediction of road pavement. By means of street pixels color exclusively, the classification accuracy reaches $85.1$\%, while including the street type attribute decreases misclassification errors and uncertainty leading to an overall accuracy of $94.5$\%. 

Despite the high accuracy of the obtained classifier, classification accuracy could be probably further improved by relying on the spatial information related to the geographical position of each road and/or its topological connectivity with adjacent roads. Road positioning is indeed available information from OpenStreetMap which was not exploited in this study. Hence, further developments should include it as input for the classification algorithm together with satellite images and street types.

The method was trained and tested on data from the Greater Maputo Area in Mozambique, but it is straightforward to predict road surfaces of all Sub-Saharan cities. Indeed, the analysis was conducted using open-source software and free datasets, and both data pre-processing and algorithm implementation phases developed advanced and scalable data science algorithms to automate the classification procedure. Future studies should replicate the proposed procedure with other case studies beyond Maputo. As a result, accurate mappings of predicted road pavement surfaces would be available in developing countries and, coherently with the mission of the Safari Njema project,
efficient solutions would be developed from them for redesigning current mobility systems.

\begin{appendix}

\section{Cluster size reduction} \label{app:N-choice}

Fundamental for $k$-NN classification is the computation of distances between street pixel clusters. The proposed measures are all based on the ground Euclidean distance between single pixels (see appendix \ref{app:distances}), whose computation is highly time consuming due to the large number of dense pixel clusters. Given a new image to be classified, the computation of distances has a time complexity of $O(Nn^2p)$, where $p$ is the dimension of the space and $N$ the number of training pixel clusters of size $n$. Since $p=3$ is fixed by the shape of the color space, and $N=2\,558$ represents the number of roads of known pavement, the reduction of the time complexity of the method must involve $n$, the size of pixel clusters.  

$n$ differs for each image and ranges between $150$ and over $5\,500$ pixels. If we suppose n to be equal to the mean, $n=2\,000$, an extremely high time complexity is obtained, namely, the classification of a new observation would use $34.48$ minutes just to compute distances, and thus is not feasible in practical applications. Note that mean computational times of label prediction are referred to the $k$-NN classification algorithm with Energy distance in the RGB space implemented in R.

The value of $n$ should be chosen not to modify the geometric structure of pixel distributions, and so not to affect the object-oriented classification. By graphical inspection, $n\!\geq\!100$ was proven to adequately represent the full clouds of pixels in the RGB space. Furthermore, distances between pixel distributions are properly approximated when $n\!\geq\!150$.  With the choice of $n=150$, the computational time of classification significantly improves, with $13.50$ seconds for each new observation to be classified. 

\section{Distances between distributions} \label{app:distances}

The object-oriented approach requires the definition of the distance between street pixel objects. They can be studied as random samples from multivariate distributions in the RGB color space and, thus, the measure of their distance is done through the computation of distances between probability measures.  

The next paragraphs provide a brief formal description of the four distances considered.

\paragraph{Euclidean distance between Tukey medians}
    This approach is inspired by the centroid linkage criterion in agglomerative hierarchical clustering. Each street cluster was represented by its single deepest point, the Tukey median \citep[see][]{Tukey1974, Rousseeuw1998}. The distance between Tukey medians was then measured with Euclidean distance in the RGB space. Medians were computed by means of the R package \texttt{DepthProc} \citep{pack-DP}.

\paragraph{Hausdorff distance}
    It is defined as the supremum of the infimum distance between two convex sets of a metric space. 
\begin{equation*}
	d_{h}(A,B)  = \max\left\{ \sup_{a\in A} \inf_{b\in B} d(a,b),\:\sup_{b\in B} \inf_{a\in A} d(a,b)\right\}
\end{equation*}
where $A, B$ are subsets of a metric space $(M,d)$. The Hausdorff distance measures the extent to which each point of a set lies near some point of another set and vice-versa \citep[see][]{Huttenlocher1993}.  

\paragraph{Energy distance}
    Energy distance was introduced by \citet{Szekely2004} as a metric for testing the equality of multivariate distributions. It is defined as follows:

\small{
\begin{equation*}
	d_{e}(A,B) = \tfrac{n_a n_b}{n_a+n_b} \left[ \tfrac{2}{n_a n_b} \sum_{a\in A,b \in B} d(a,b) 
	-  \tfrac{1}{n_a^2}\sum_{a,a' \in A} d(a,a') - \tfrac{1}{n_b^2} \sum_{b,b'\in B} d(b,b') \right]
\end{equation*}}

where $A, B$ are finite subsets of a metric space $(M,d)$, $|A|=n_a$, $|B|=n_b$. Roughly speaking, it compares the distance of points from different distributions with the one within the same distribution. The R package \texttt{energy} \citep{pack-energy} was used to compute such distance in an optimized way.

\paragraph{Wasserstein distance} 
It measures the minimal amount of work that must be performed to transform one distribution into another by moving “distribution mass” around. Its computation is strictly related to the resolution of a transport problem; indeed, the R package \texttt{transport} \citep{pack-transport} was used to compute Wasserstein distances. 

Let $\mu, \nu$ be two probability distributions in a metric space $(M,d)$, denote with $\Gamma(\mu,\nu)$ the space of measures with marginals $\mu$ and $\nu$, and let $c(\cdot,\cdot)$ be any cost, such as d$(\cdot,\cdot)$. Then, the Wasserstein distance is defined as follows:
\begin{equation*}
    d_{w}(\mu, \nu)  = \inf_{\gamma \in \Gamma(\mu, \nu)} \int_{M\times M} c(x,y) \: d\gamma(x,y)
\end{equation*}
For a detailed description, see \citet{Panaretos2019}.

In previous formulas, $d$ represents the ground distance and was chosen to be the Euclidean distance in RGB space: \mbox{$d(a,b) = \sqrt{(R_a - R_b)^2+ (G_a - G_b)^2 + (B_a - B_b)^2}$.}

As discussed in section \ref{sec:analysis}, when using Tukey median distance, the information of the whole cloud of points is summarized into one of its pixels with which the classification is performed. The method is hence a borderline approach of object-oriented classification, as an extreme case of distance between clouds. We decided to take it into consideration for two reasons. Firstly, the distribution of Tukey medians in the RGB space suggested the existence of two separate clusters. The group of points on the diagonal was mostly composed by medians of paved roads, while medians of unpaved roads were distributed below the diagonal. Then, although the computation of Tukey medians of training data is highly time consuming, the classification of new instances is faster than the one with other distances. However, eventually the $k$-NN classifier with other distances is more accurate, making almost half of mistakes, and has a still acceptable time complexity (see table \ref{tab:metrics}), so was preferred.

\section{Color Spaces} \label{app:colors}

Color models are systems which provide a rational method for measuring colors. The selection of a color space should consider the application field of the analysis since different models are suitable to address different problems. For this reason, $k$-NN algorithm was trained to classify data from three different color spaces, namely RGB, R’G’B’ and HSV. This study refers to the work of \citet{Plataniotis2013}, which collects valuable and detailed descriptions of color spaces. Here we give an overview of their definition, advantages and disadvantages.

\paragraph{RGB space}
    It is a $3$-dimensional space whose components represent the three color bands of red, green and blue. Values of each component ranges from $0$ to $255$, shaping the whole space as a cube. The RGB system is additive and computationally practical and is considered as the base color model for most image applications. However, the interpretation of its components is hard because far from human perception mechanisms. Humans interpret colors based on lightness, saturation, and hue, rather than as a sum of red, green and blue elements. So, the RGB space is not suitable for the analysis of perceptual attributes.

\paragraph{R’G’B’ space}
    It is a non-linear transformation of the RGB space, formally described in the formula below. RGB values must be rescaled in the range $[0,1]$ before applying the transformation to each component $C\! \in\! \{R, G, B\}$, and values $C^\prime \! \in\! \{R^\prime, G^\prime, B^\prime\}$ should be readjusted back in the range $[0,255]$.

\begin{equation*}
	C^\prime = \begin{cases}
		4.5\: C, & \text{if } C \leq 0.018 \\
		1.099\: C ^{- \gamma}\: -0.099, & \text{otherwise} 
	\end{cases}
\label{eq:R'G'B'}
\end{equation*}

The gamma correction parameter $\gamma$ assumes different values depending on the application system. Acceptable values for $\gamma$ range between 2.2 and 2.6. For this study, $\gamma=2.2$. R'G’B’ introduces non-linearity to mimic the human visual system, non-linear with respect to color intensity perception. R'G'B' values are less perceptually non-uniform than linear RGB, but still not adequately uniform and accurate for perceptual computations.

\paragraph{HSV space}
    HSV components collect information on hue H, saturation S and value of brightness V. This color system uses approximately cylindrical coordinates: H $\in [0,2\pi]$ is proportional to the angle, S $\in [0,1]$ to the radial distance, and V $\in [0,1]$ to the height along the axis. 
    The transformation between RGB and HSV is highly non-linear and can be found in \citet{Plataniotis2013}. Just to have an idea on the relationship between RGB and HSV, one can imagine that the RGB diagonal coincides with the HSV height. Consequently, saturation is related to the distance from the diagonal, hue to the angle of the projection on the diagonal, and value quantifies the distance from the black vertex in both cases.  

Street pixel transformation from RGB to HSV was performed by means of the function \texttt{rgb2hsv} in the R package \texttt{grDevices}. Data analysis was then performed in the hexcone model which uses chroma rather than saturation ($\text{chroma}\!=\!\text{SV}$ is the saturation relative to a certain value). It is indeed more suitable for the intuitive notion of color purity.
The HSV model was built to comply with the human perception of colors and therefore is intuitive and easily interpretable. A relevant issue of the HSV system is the instability of hue for low saturated points which often occurred for street pixels in this study. 

The $k$-NN classification method was performed in different color spaces using Energy distance between distributions. The Cartesian distance was used as ground distance of Energy metric in each of them, specifically L\textsubscript{2} distance in the RGB and R’G’B’ cubic spaces, while in the HSV hexcone space
\\
\mbox{$\text{d}(a,b)= \sqrt{\Delta(SV \sin H)^2 + \Delta(SV \cos H)^2 + \Delta V^2}$}. Distances in RGB and R’G’B’ were computed in R, while distances in HSV were implemented in C\texttt{++} to speed up computations.

\section{Classification with images and street type} \label{app:type}

\begin{figure}[tb]
    \centering
    \includegraphics[scale=0.5]{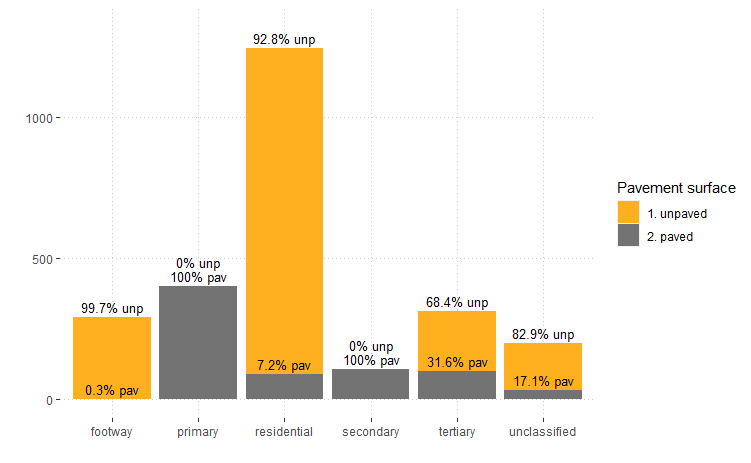}
    \caption{Distribution of paved and unpaved labels among street types}
    \label{fig:barplot}
\end{figure}

The type of a street and its pavement surface are expected to be related attributes. The intuition is supported by the distribution of paved and unpaved roads within different street types. Primary and secondary roadways of known pavement are always paved, unlike footway paths which are almost always unpaved (see figure \ref{fig:barplot}). Moreover, the street type attribute is very informative for road surface classification. The simple road surface classification rule which only depends on type, namely assigns a pavement class as the most frequent one within the group of a road type, has a MER of $0.087$ over the test set.

In view of the above, the proposed classification algorithm was expanded to consider not only satellite image data, but also the street type attribute, when available. The main scheme of the classification algorithm remains essentially the same. The difference is in the research of neighbors, where only instances of the same type are considered. It might be said that six classifiers, one for each street type, work in parallel to predict the road pavement surface and only cooperate during the parameter tuning phase. Indeed, $k=5$ was set to minimize the total misclassification error rate of 10-fold cross validation, and $f_{up} =0.6$ minimizes the total cost of the method, summing errors and costs of each classifier. The classification algorithm results in a MER of $0.0548$ and confusion matrix reported in table \ref{tab:confusion2}.

\begin{table}[tb]
    \centering
    \caption{Confusion matrix of $k$-NN classifier from satellite images and street type attribute over the test set, with $k=5$ and $f_{up} =0.6$} 
    \label{tab:confusion2}
    \begin{tabular}{p{1cm}|p{1.5cm}| p{1.5cm}| p{1.5cm}|}
		
		\multicolumn{2}{c}{}&\multicolumn{2}{c}{\textit{Predicted}}\\
		\cline{3-4}
		\multicolumn{2}{c|}{}&paved&unpaved\\
		\cline{2-4}
		\textit{True} &paved & $186$ & $28$ \\
		\cline{2-4} &unpaved & $14$ & $539$ \\
		\cline{2-4}
    \end{tabular}
\end{table}

An attempt was made to build six fully independent classifiers with different optimal parameters $k$ and $f$. Results are collected in table \ref{tab:dif-k-types}. The global performance slightly improves ($\text{MER}=0.0521$) with respect to the most general method. Since the difference is small, it can be asserted that the more general classification method with common parameters is a good approximation of type-specific case and, as simpler, is preferred.

\begin{table}[tbh]
    \centering
    \caption{Optimal number of neighbors $k$ and threshold $f$ for independent classifiers of different street types, resulting in different numbers of mistakes {N\textsubscript{pu}} and {N\textsubscript{up}}, and uncertainty {N\textsubscript{d}}. Symbol ``-" is used when each plausible value of the parameter is optimal
    }
    \label{tab:dif-k-types}

    \begin{tabular}{l@{\hskip 50pt}cc@{\hskip 35pt}ccc}
        \hline
        {Type} & {k} & {$f$} & {N\textsubscript{up}} & {N\textsubscript{pu}} & {N\textsubscript{d}}\\
        \hline
        {Primary}       & - & -                        & 0  & 0    & 0  \\
        {Secondary}     & - & -                        & 0  & 0    & 0 \\
        {Tertiary}      & 7 & $f_{up}=0.57$ & 10 & 5    & 0  \\
        {Unclassified}  & 7 & $f_{up}=0.57$ & 1  & 8    & 0 \\
        {Residential}   & 5 & $f_{up}=0.4$  & 5  & 10   & 0 \\
        {Footways}      & - & -                        & 0  & 0    & 0 \\
        \hline 
    \end{tabular}

\end{table}

OSM streets type are more than $40$, each one with different importance and purpose in the road network \citep{wiki-OSM}. Uncommon types were merged with the principal $6$ tags, asserting that there was no need for highly specific type descriptions for pavement surface classification. For the same reason, other super-categories of street type were considered. The easiest and most relevant categorization unifies primary with secondary roads, as well as tertiary with unclassified, and finally residential with footways. Combinations were chosen due to image similarity shown during graphical inspection, and to the importance level and functioning of such roads. Applying the extended $k$-NN classifier with super-categories of street types results in a MER of $0.059$, very similar to the performance of the algorithm which uses $6$ actual street types. The result suggests that the division between types is not strict for the purpose of pavement analysis and supports type reduction to $6$ types.

\end{appendix}

\bibliographystyle{abbrvnat}
\bibliography{references}

\begin{thebibliography}{32}
\providecommand{\natexlab}[1]{#1}
\providecommand{\url}[1]{\texttt{#1}}
\expandafter\ifx\csname urlstyle\endcsname\relax
  \providecommand{\doi}[1]{doi: #1}\else
  \providecommand{\doi}{doi: \begingroup \urlstyle{rm}\Url}\fi

\bibitem[Abdollahi et~al.(2020)Abdollahi, Pradhan, Shukla, Chakraborty, and Alamri]{Abdollahi2020}
A.~Abdollahi, B.~Pradhan, N.~Shukla, S.~Chakraborty, and A.~Alamri.
\newblock Deep learning approaches applied to remote sensing datasets for road extraction: A state-of-the-art review.
\newblock \emph{Remote Sensing}, 12\penalty0 (9), 2020.
\newblock ISSN 2072--4292.
\newblock \doi{10.3390/rs12091444}.

\bibitem[Behrens et~al.(2016)Behrens, McCormick, and Mfinanga]{Behrens2016}
R.~Behrens, D.~McCormick, and D.~Mfinanga.
\newblock \emph{{Paratransit in African Cities: Operations, Regulation and Reform}}.
\newblock Routledge, 2016.

\bibitem[Chow(1970)]{Chow1970}
C.~Chow.
\newblock On optimum recognition error and reject tradeoff.
\newblock \emph{IEEE Transactions on Information Theory}, 16\penalty0 (1):\penalty0 41--46, 1970.
\newblock \doi{10.1109/TIT.1970.1054406}.

\bibitem[Ester et~al.(1996)Ester, Kriegel, Sander, and Xu]{Ester1996}
M.~Ester, H.-P. Kriegel, J.~Sander, and X.~Xu.
\newblock {A Density-Based Algorithm for Discovering Clusters in Large Spatial Databases with Noise}.
\newblock In \emph{KDD}, volume~96, pages 226--231. Association for the Advancement of Artificial Intelligence AIII Press, 1996.

\bibitem[{Google Earth}(2020)]{google-earth-imagery}
{Google Earth}, 2020.
\newblock URL \url{http://www.google.com/earth/index.html}.
\newblock Accessed: 2020/04/20.

\bibitem[Hennig(2020)]{pack-fpc}
C.~Hennig.
\newblock \emph{fpc: Flexible Procedures for Clustering}, 2020.
\newblock R package version 2.2-9.

\bibitem[Hijmans(2021)]{pack-raster}
R.~J. Hijmans.
\newblock \emph{raster: Geographic Data Analysis and Modeling}, 2021.
\newblock R package version 3.4-10.

\bibitem[Hou et~al.(2016)Hou, Gao, and Li]{Hou2016}
J.~Hou, H.~Gao, and X.~Li.
\newblock {DSets-DBSCAN: A parameter-free clustering algorithm}.
\newblock \emph{IEEE Transactions on Image Processing}, 25\penalty0 (7):\penalty0 3182--3193, 2016.
\newblock \doi{10.1109/TIP.2016.2559803}.

\bibitem[Huttenlocher et~al.(1993)Huttenlocher, Klanderman, and Rucklidge]{Huttenlocher1993}
D.~P. Huttenlocher, G.~A. Klanderman, and W.~J. Rucklidge.
\newblock {Comparing images using the Hausdorff distance}.
\newblock \emph{IEEE Transactions on Pattern Analysis and Machine Intelligence}, 15\penalty0 (9):\penalty0 850--863, 1993.
\newblock \doi{10.1109/34.232073}.

\bibitem[{JICA}(2010)]{JICA2010}
{JICA}.
\newblock {O Estudo Preparat\'{o}rio sobre Plano de Melhoramenro da Estrada no Corredor de Desenvolvimento de Nacala (N13: Cuamba-Mandimba-Lichinga) na Rep\'{u}blica de Mo\c{c}ambique}.
\newblock Final report, Administra\c{c}\~{a}o Nacional de Estradas - Rep\'{u}blica de Mo\c{c}ambique, 2010.

\bibitem[Kosiorowski and Zawadzki(2020)]{pack-DP}
D.~Kosiorowski and Z.~Zawadzki.
\newblock \emph{DepthProc: Statistical Depth Functions for Multivariate Analysis}, 2020.
\newblock R package version 2.1.3.

\bibitem[Kurumalla and Rao(2016)]{Kurmalla2016}
S.~Kurumalla and P.~S. Rao.
\newblock {K-nearest neighbor based DBSCAN clustering algorithm for image segmentation}.
\newblock \emph{Journal of Theoretical and Applied Information Technology}, 92\penalty0 (2):\penalty0 395--402, 10 2016.

\bibitem[Li et~al.(2019)Li, Chen, Zhao, and Chen]{Li2019}
C.~Li, S.~Chen, Y.~Zhao, and Y.~Chen.
\newblock {Road Pavement Identification based on Acceleration Signals of Off-road Vehicles Using the Batch Normalized Recurrent Neural Network}.
\newblock In \emph{IEEE International Conference on Artificial Intelligence and Computer Applications}, pages 172--177, 2019.
\newblock \doi{10.1109/ICAICA.2019.8873458}.

\bibitem[Marianingsih et~al.(2019)Marianingsih, Utaminingrum, and Bachtiar]{Marianingsih2019}
S.~Marianingsih, F.~Utaminingrum, and F.~A. Bachtiar.
\newblock {Road Surface types classification using combination of K-nearest neighbor and Naïve Bayes based on GLCM}.
\newblock \emph{International Journal of Advances in Soft Computing and its Applications}, 11\penalty0 (2):\penalty0 15--27, 2019.

\bibitem[Marron and Alonso(2014)]{Marron2014}
J.~S. Marron and A.~M. Alonso.
\newblock Overview of object oriented data analysis.
\newblock \emph{Biometrical Journal}, 56\penalty0 (5):\penalty0 732--753, 2014.
\newblock \doi{https://doi.org/10.1002/bimj.201300072}.

\bibitem[Marron and Dryden(2021)]{Marron2021}
J.~S. Marron and I.~L. Dryden.
\newblock \emph{{Object Oriented Data Analysis}}.
\newblock Chapman and Hall/CRC, 1st edition, 2021.
\newblock ISBN 9781351189675.

\bibitem[{OpenStreetMap contributors}(2020)]{OSM}
{OpenStreetMap contributors}.
\newblock {Planet dump retrieved from \url{https://planet.osm.org }}, 2020.
\newblock URL \url{https://www.openstreetmap.org}.

\bibitem[{OpenStreetMap Wiki}(2021)]{wiki-OSM}
{OpenStreetMap Wiki}, 2021.
\newblock URL \url{https://wiki.openstreetmap.org}.
\newblock Accessed: 2021/08/21.

\bibitem[Panaretos and Zemel(2019)]{Panaretos2019}
V.~M. Panaretos and Y.~Zemel.
\newblock {Statistical aspects of Wasserstein distances}.
\newblock \emph{Annual review of statistics and its application}, 6:\penalty0 405--431, 2019.
\newblock \doi{10.1146/annurev-statistics-030718-104938}.

\bibitem[Paulo et~al.(2010)Paulo, Coelho, and Figueiredo]{Paulo2018}
J.~P. Paulo, J.~L.~B. Coelho, and M.~A.~T. Figueiredo.
\newblock Statistical classification of road pavements using near field vehicle rolling noise measurements.
\newblock \emph{The Journal of the Acoustical Society of America}, 128\penalty0 (4):\penalty0 1747--1754, 2010.
\newblock \doi{10.1121/1.3466870}.

\bibitem[Pillai et~al.(2013)Pillai, Fumera, and Roli]{Pillai2013}
I.~Pillai, G.~Fumera, and F.~Roli.
\newblock Multi-label classification with a reject option.
\newblock \emph{Pattern Recognition}, 46\penalty0 (8):\penalty0 2256--2266, 2013.
\newblock \doi{https://doi.org/10.1016/j.patcog.2013.01.035}.

\bibitem[Plataniotis and Venetsanopoulos(2013)]{Plataniotis2013}
K.~N. Plataniotis and A.~N. Venetsanopoulos.
\newblock \emph{Color image processing and applications}, chapter Color spaces.
\newblock Springer Science \& Business Media, 2013.

\bibitem[Ragnoli et~al.(2018)Ragnoli, De~Blasiis, and Di~Benedetto]{Ragnoli2018}
A.~Ragnoli, M.~R. De~Blasiis, and A.~Di~Benedetto.
\newblock {Pavement Distress Detection Methods: A Review}.
\newblock \emph{Infrastructures}, 3\penalty0 (4), 2018.
\newblock \doi{10.3390/infrastructures3040058}.

\bibitem[Riid et~al.(2020)Riid, Manna, and Astapov]{Riid2020}
A.~Riid, D.~L. Manna, and S.~Astapov.
\newblock Image-based pavement type classification with convolutional neural networks.
\newblock In \emph{2020 IEEE 24th International Conference on Intelligent Engineering Systems}, pages 55--60, 2020.
\newblock \doi{10.1109/INES49302.2020.9147199}.

\bibitem[Rizzo and Szekely(2021)]{pack-energy}
M.~Rizzo and G.~Szekely.
\newblock \emph{energy: E-Statistics: Multivariate Inference via the Energy of Data}, 2021.
\newblock URL \url{https://CRAN.R-project.org/package=energy}.
\newblock R package version 1.7-8.

\bibitem[Rousseeuw and Ruts(1998)]{Rousseeuw1998}
P.~J. Rousseeuw and I.~Ruts.
\newblock {Constructing the bivariate Tukey median}.
\newblock \emph{Statistica Sinica}, pages 827--839, 1998.
\newblock URL \url{http://www.jstor.org/stable/24306466}.

\bibitem[{Safari Njema}(2018)]{web-SAFARI}
{Safari Njema}, 2018.
\newblock URL \url{https://www.safari-njema.polimi.it/}.
\newblock Accessed: 2021/08/30.

\bibitem[Schuhmacher et~al.(2020)Schuhmacher, Bähre, Gottschlich, Hartmann, Heinemann, and Schmitzer]{pack-transport}
D.~Schuhmacher, B.~Bähre, C.~Gottschlich, V.~Hartmann, F.~Heinemann, and B.~Schmitzer.
\newblock \emph{{transport}: Computation of Optimal Transport Plans and Wasserstein Distances}, 2020.
\newblock R package version 0.12-2.

\bibitem[Slavkovikj et~al.(2014)Slavkovikj, Verstockt, De~Neve, Hoecke, and Van~de Walle]{Slavkovikj2014}
V.~Slavkovikj, S.~Verstockt, W.~De~Neve, S.~Hoecke, and R.~Van~de Walle.
\newblock {Image-Based Road Type Classification}.
\newblock In \emph{IEEE International Conference on Pattern Recognition}, pages 2359--2364, 2014.
\newblock \doi{10.1109/ICPR.2014.409}.

\bibitem[Sz{\'e}kely and Rizzo(2004)]{Szekely2004}
G.~J. Sz{\'e}kely and M.~L. Rizzo.
\newblock Testing for equal distributions in high dimension.
\newblock \emph{InterStat}, 5\penalty0 (16.10):\penalty0 1249--1272, 2004.

\bibitem[Tsunokawa and Changyu(2002)]{Tsunokawa2002}
K.~Tsunokawa and G.~Changyu.
\newblock Optimal strategies for highway pavement management in developing countries.
\newblock \emph{Computer-Aided Civil and Infrastructure Engineering}, 17\penalty0 (3):\penalty0 194--202, 2002.

\bibitem[Tukey(1974)]{Tukey1974}
J.~W. Tukey.
\newblock {Mathematics and the Picturing of Data}.
\newblock In \emph{Proceedings of the International Congress of Mathematicians held in Vancouver}, volume~2. Canadian Mathematical Congress, 1974.

\end{thebibliography}

\end{document}